\documentclass[10pt,preprintnumbers,showpacs]{revtex4}
\usepackage{graphicx,epsfig,axodraw}

\def\xslash#1{{\rlap{$#1$}/}}
\def\half{\frac{1}{2}}
\def\beq{\begin{equation}}
\def\eeq{\end{equation}}
\def\beqa{\begin{eqnarray}}
\def\eeqa{\end{eqnarray}}
\def\iar{\begin{array}{l}}
\def\ear{\end{array}}

\begin{document}

\title{Unstable particle's wave-function renormalization prescription}
\author{Yong Zhou}
\affiliation{Institute of High Energy Physics, Academia Sinica, P.O. Box 918(4), Beijing 100049, China, Email: zhouy@ihep.ac.cn}

\begin{abstract}
We strictly define two set Wave-function Renormalization Constants (WRC) under the LSZ reduction formula for unstable particles at the first time. Then by introducing antiparticle's WRC and the CPT conservation law we obtain a new wave-function renormalization condition which can be used to totally determine the two set WRC. We calculate two physical processes to manifest the consistence of the present wave-function renormalization prescription with the gauge theory in standard model. We also prove that the conventional wave-function renormalization prescription which discards the imaginary part of unstable particle's WRC leads to physical amplitude gauge dependent.
\end{abstract}

\maketitle

\section{Introduction}

Wave-function or field renormalization prescriptions have been present for a long time, but at present they encounter some problems for unstable particles \cite{c1,c2,c3}. Since there is imaginary part present in unstable particle's self energy, how to deal with it in determining unstable particle's WRC becomes an inevitable problem. We begin our discussion from the fermion field renormalization prescriptions. The fermion Field Renormalization Constants (FRC) (which serve as fermion WRC) can be introduced as \cite{c2}
\beq
  \Psi_{0i}\,=\,\sum_j Z^{\half}_{ij}\Psi_j\,, \hspace{10mm}
  \bar{\Psi}_{0i}\,=\,\sum_j \bar{\Psi}_j\bar{Z}^{\half}_{ji}\,,
\eeq
with
\beq 
  Z^{\half}_{ij}\,=\,Z^{L\half}_{ij}\gamma_L + Z^{R\half}_{ij}\gamma_R\,,\hspace{10mm}
  \bar{Z}^{\half}_{ij}\,=\,\bar{Z}^{L\half}_{ij}\gamma_{R} +
  \bar{Z}^{R\half}_{ij}\gamma_{L}\,, 
\eeq
where $i$ and $j$ are the fermion generations, and $\gamma_L$ and $\gamma_R$ are the left- and right- handed helicity operators. Because the bare fermion fields have the relationship $\bar{\Psi}_{0i}=\Psi_{0i}^{\dagger}\gamma^0$, the fermion FRC seem to satisfy the `pseudo-hermiticity' relationship \cite{c2,c3}
\beq
  \bar{Z}^{\half}_{ij}\,=\,\gamma^{0}Z^{\half\dagger}_{ij}\gamma^{0}\,.
\eeq
It's well known that the conventional field renormalization prescription for fermions is
\beqa
  \hat{\Gamma}_{ij}(p)\,u_j(p)|_{p^2=m_j^2}\,=\,0\hspace{-3mm}&&, \hspace{10mm}
  \bar{u}_{i}(p)\hat{\Gamma}_{ij}(p)|_{p^2=m_i^2}\,=\,0\,, \nonumber \\
  \lim_{p^2\rightarrow m_i^2}\frac{{\xslash p}+m_i}{p^2-m_i^2}\hat{\Gamma}_{ii}(p)\,
  u_i(p)\,=\,u_i(p)\hspace{-3mm}&&, \hspace{10mm}
  \lim_{p^2\rightarrow m_i^2}\bar{u}_i(p)\hat{\Gamma}_{ii}(p)
  \frac{{\xslash p}+m_i}{p^2-m_i^2}\,=\,\bar{u}_i(p)\,,
\eeqa
where $\hat{\Gamma}_{ij}$ is the renormalized fermion two-point function. Of course there is no problem for stable fermions, but for unstable fermions Eqs.(4) must be revised since the imaginary part coming from the branch cut of the fermion self energy makes Eq.(3) and Eqs.(4) cannot be simultaneously satisfied \cite{c2,c3}. The acceptable fermion field renormalization prescription under the constraint of Eq.(3) should be \cite{c4}
\beqa
  \tilde{Re}\,\hat{\Gamma}_{ij}(p)\,u_j(p)|_{p^2=m_j^2}\,=\,0\hspace{-3mm}&&, 
  \hspace{10mm}
  \tilde{Re}\,\bar{u}_{i}(p)\hat{\Gamma}_{ij}(p)|_{p^2=m_i^2}\,=\,0\,, \nonumber \\
  \lim_{p^2\rightarrow m_i^2}\frac{{\xslash p}+m_i}{p^2-m_i^2}\tilde{Re}\,
  \hat{\Gamma}_{ii}(p)\,u_i(p)\,=\,u_i(p)\hspace{-3mm}&&, \hspace{10mm}
  \lim_{p^2\rightarrow m_i^2}\bar{u}_i(p)\tilde{Re}\,\hat{\Gamma}_{ii}(p)
  \frac{{\xslash p}+m_i}{p^2-m_i^2}\,=\,\bar{u}_i(p)\,,
\eeqa
where $\tilde{Re}$ takes the left part of the self energy after removing the branch cut of it.

But such fermion field renormalization prescription makes physical amplitude gauge-parameter dependent (see Ref.\cite{c2,c5} and the discussion below). The only way to solve this problem is to discard the constraint of Eq.(3) for unstable external-line fermions of S-matrix, since Eq.(3) has been broken by the branch cut of unstable fermion's self energy \cite{c2}. Under this prescription for diagonal fermion FRC D. Espriu et al. obtain \cite{c2}
\beqa
  \delta\bar{Z}^L_{ii}&=&-\Sigma^L_{ii}(m_i^2)-m_i^2
  \frac{\partial}{\partial p^2}\left[\Sigma^L_{ii}+
  \Sigma^R_{ii}+2\Sigma^S_{ii} \right]_{p^2=m_i^2}-\frac{\alpha_i}{2}\,,
  \nonumber \\
  \delta\bar{Z}^R_{ii}&=&-\Sigma^R_{ii}(m_i^2)-m_i^2
  \frac{\partial}{\partial p^2}\left[\Sigma^L_{ii}+
  \Sigma^R_{ii}+2\Sigma^S_{ii} \right]_{p^2=m_i^2}-\frac{\alpha_i}{2}\,,
  \nonumber \\
  \delta Z^L_{ii}&=&-\Sigma^L_{ii}(m_i^2)-m_i^2
  \frac{\partial}{\partial p^2}\left[ \Sigma^L_{ii}+
  \Sigma^R_{ii}+2\Sigma^S_{ii} \right]_{p^2=m_i^2}+\frac{\alpha_i}{2}\,,
  \nonumber \\
  \delta Z^R_{ii}&=&-\Sigma^R_{ii}(m_i^2)-m_i^2
  \frac{\partial}{\partial p^2}\left[ \Sigma^L_{ii}+
  \Sigma^R_{ii}+2\Sigma^S_{ii} \right]_{p^2=m_i^2}+\frac{\alpha_i}{2}\,,
\eeqa
where the one-loop fermion self energy is written as
\beq
  \Sigma_{ii}(\xslash p)\,=\,\xslash p\gamma_L\Sigma^L_{ii}(p^2)+\xslash p\gamma_R
  \Sigma^R_{ii}(p^2)+m_i\Sigma^{S}_{ii}(p^2)\,,
\eeq
and $\alpha_i$ is an arbitrary coefficient. Since there are arbitrary coefficients in Eqs.(6), the definition of fermion FRC has indetermination. This indetermination doesn't affect the neutral current couplings at one-loop level, but changes the charged current couplings. Generally speaking this indetermination cannot be removed in physical results, for example in the physical result of $W^{-}\rightarrow e^{-}\bar{\nu}_e$.

Besides the indetermination, there is also an unclear problem in this renormalization prescription. There are two set FRC $Z^{\half}_{ij}$ and $\bar{Z}^{\half}_{ij}$ respectively for the incoming and outgoing fermions, but how to determine the incoming and outgoing anti-fermion's FRC hasn't been clearly discussed. In this paper we introduce four set WRC, two set for particles and two set for antiparticles, and discuss how to totally determine them. The arrangement of the contents is as follows. We firstly strictly define the particle's WRC under the LSZ reduction formula \cite{c6}. Then we discuss how to introduce the antiparticle's WRC and use them together with the CPT conservation law to obtain a new wave-function renormalization condition to totally determine the WRC. In section 4 we calculate two physical processes to manifest that compared with the conventional one the present wave-function renormalization prescription keeps physical amplitude gauge invariant. Lastly we give our conclusion.

\section{Definition of wave-function renormalization constants under the LSZ reduction formula}

Generally speaking there are two ways to determine the WRC: one is to introduce FRC which serve as WRC \cite{c1,c2,c3,c4}, the other is not to introduce FRC and the WRC are determined by the LSZ reduction formula. At present we only discuss the second prescription.

From the LSZ reduction formula the WRC is the {\em field strength renormalization factor}. For boson one has \cite{c6}
\beq
  \int d^4 x\,e^{i\,p_j\cdot x}\cdot\cdot\cdot\int d^4 y\,e^{-i\,p_i\cdot y}
  <\hspace{-1mm}\Omega|T\{\phi_j(x)\cdot\cdot\cdot\phi_i^{\dagger}(y)\}|
  \Omega\hspace{-1mm}>\,\sim\,
  \frac{i<\hspace{-1mm}\Omega|\phi_j(0)|\lambda_j\hspace{-1mm}>}{p_j^2-m_j^2+i\epsilon}
  \cdot\cdot\cdot \frac{i<\hspace{-1mm}\lambda_i|\phi_i^{\dagger}(0)|\Omega\hspace{-1mm}>}
  {p_i^2-m_i^2+i\epsilon}<\hspace{-1mm}j\cdot\cdot\cdot|S|\cdot\cdot\cdot i\hspace{-1mm}>\,,
\eeq
where $\Omega$ is the interaction vacuum, $T$ is the time-ordering operator, $\phi_i$ and $\phi_j$ are the Heisenberg fields, $p_i$ and $p_j$ are on mass shell, and $\lambda_i$ and $\lambda_j$ are the incoming and outgoing boson states of S-matrix elements. We can introduce boson WRC as follows
\beq
  Z_i^{\half}\,=\,<\hspace{-1mm}\Omega|\phi_i(0)|\lambda_i\hspace{-1mm}>\,, \hspace{10mm}
  \bar{Z}_i^{\half}\,=\,<\hspace{-1mm}\lambda_i|\phi_i^{\dagger}(0)|\Omega\hspace{-1mm}>
  \,.
\eeq
We note that the LSZ reduction formula has only been proved for stable particles \cite{c6}. Here we will postulate a generalization of the LSZ reduction formula to unstable particles. Under the postulation the following formulas and Eqs.(8,9) also hold true for unstable particles. Note that the $\epsilon$ in Eq.(8) isn't infinitesimal any more for unstable particles under the postulation, since it will be proportional to the unstable particle's decay width according to the Breit-Wigner formula \cite{cin}. From Eqs.(8,9) the the scalar boson's propagation amplitude in interaction vacuum is
\beq
  \int d^4 x\,e^{i\,p\cdot x}<\hspace{-1mm}\Omega|T\{\phi_i(x)\phi_i^{\dagger}(0)\}|
  \Omega\hspace{-1mm}>\,\sim\,
  \frac{i<\hspace{-1mm}\Omega|\phi_i(0)|\lambda_i\hspace{-1mm}>
  <\hspace{-1mm}\lambda_i|\phi_i^{\dagger}(0)|\Omega\hspace{-1mm}>}{p^2-m_i^2+i\epsilon}
  \,=\,\frac{i\,Z_i^{\half}\bar{Z}_i^{\half}}{p^2-m_i^2+i\epsilon}\,.
\eeq
Form Eq.(10), Eq.(8) can also be written as
\beq
  \int d^4 x\,e^{i\,p_j\cdot x}\cdot\cdot\cdot\int d^4 y\,e^{-i\,p_i\cdot y}
  <\hspace{-1mm}\Omega|T\{\phi_j(x)\cdot\cdot\cdot\phi_i^{\dagger}(y)\}|
  \Omega\hspace{-1mm}>\,\sim\,
  \frac{i\,Z_j^{\half}\bar{Z}_j^{\half}}{p_j^2-m_j^2+i\epsilon}
  \cdot\cdot\cdot \frac{i\,Z_i^{\half}\bar{Z}_i^{\half}}{p_i^2-m_i^2+i\epsilon}
  {\cal M}^{amp}(i\cdot\cdot\cdot\rightarrow \cdot\cdot\cdot j)\,,
\eeq
where the superscript $amp$ represents the amputated Feynman amplitude. From Eqs.(8,9,11) we obtain the familiar result
\beq
  <\hspace{-1mm}j\cdot\cdot\cdot|S|\cdot\cdot\cdot i\hspace{-1mm}>\,=\,\bar{Z}_j^{\half}
  {\cal M}^{amp}(i\cdot\cdot\cdot\rightarrow \cdot\cdot\cdot j)Z_i^{\half}\,,
\eeq
where the other external-line particle's WRC are ignored for convenience. Thus $Z_i^{\half}$ is the incoming boson's WRC, and $\bar{Z}_j^{\half}$ is the outgoing boson's WRC. 

From Eq.(10) the boson's WRC can be obtained by expanding the boson propagation amplitude at $p^2\rightarrow m_i^2$ 
\beq
  \frac{i}{p^2-m_i^2-\delta m_i^2+\Sigma_{ii}(p^2)}\,\sim\,
  \frac{i}{(p^2-m_i^2)(1+\Sigma_{ii}^{\prime}(m_i^2))+\Sigma_{ii}(m_i^2)-\delta m_i^2}
  \,=\,\frac{i\,(1+\Sigma_{ii}^{\prime}(m_i^2))^{-1}}{p^2-m_i^2+i\,\epsilon}\,,
\eeq
where $\Sigma_{ii}^{\prime}(m_i^2)=\partial\Sigma_{ii}(m_i^2)/\partial p^2$, and $\epsilon=(1+\Sigma_{ii}^{\prime}(m_i^2))^{-1}(Im\,\Sigma_{ii}(m_i^2)-i\,Re\, \Sigma_{ii}(m_i^2)+i\,\delta m_i^2)$ is a small quantity which is proportional to the boson's decay width at one-loop level. Therefore one has from Eqs.(10,13)
\beq
  Z_i^{\half}\bar{Z}_i^{\half}\,=\,(1+\Sigma_{ii}^{\prime}(m_i^2))^{-1}\,.
\eeq
On the other hand, from the relationship between the two matrix elements: $<\hspace{-1mm}\lambda_i|\phi_i^{\dagger}(0)|\Omega\hspace{-1mm}>\,=\,<\hspace{-1mm}\Omega|\phi_i(0)|\lambda_i\hspace{-1mm}>^{\dagger}$, we obtain a hermitian conjugation relationship between the incoming and outgoing boson's WRC (see Eqs.(9))
\beq
  \bar{Z}_i^{\half}\,=\,Z_i^{\half\ast}\,.
\eeq
But such hermitian conjugation relationship has been broken by the imaginary parts of unstable boson's propagation amplitudes \cite{c5}. This can also been seen from Eq.(14): Eq.(15) requires Eq.(14) must be a real number, but it is obviously not the case for unstable bosons. 

For fermion the above discussions become a little more complex. We note that there has been an early discussion of this problem in Ref.\cite{c8}, in which part of the conclusion will be cited in the following discussions. For the Green function with two fermion external lines we have (like Eq.(8))
\beq
  \int d^4 x\,e^{i\,p_j\cdot x}\cdot\cdot\cdot\int d^4 y\,e^{-i\,p_i\cdot y}
  <\hspace{-1mm}\Omega|T\{\psi_j(x)\cdot\cdot\cdot\bar{\psi}_i(y)\}|
  \Omega\hspace{-1mm}>\,\sim\,
  \frac{i<\hspace{-1mm}\Omega|\psi_j(0)|\lambda_{j_s}\hspace{-1mm}>}
  {p_j^2-m_j^2+i\epsilon}<\hspace{-1mm}j_s\cdot\cdot\cdot|S|
  \cdot\cdot\cdot i_s\hspace{-1mm}>\frac{i<\hspace{-1mm}\lambda_{i_s}|
  \bar{\psi}_i(0)|\Omega\hspace{-1mm}>}{p_i^2-m_i^2+i\epsilon}\,,
\eeq
where the subscript $s$ denotes the helicity of the fermion states, and the other boson WRC and propagators have been ignored for convenience. The fermion WRC can be introduced as follows:
\beq
  <\hspace{-1mm}\Omega|\psi_i(0)|\lambda_{i_s}\hspace{-1mm}>\,=\,Z_i^{\half}u_i^s\,,
  \hspace{10mm}
  <\hspace{-1mm}\lambda_{i_s}|\bar{\psi}_i(0)|\Omega\hspace{-1mm}>\,=\,\bar{u}_i^s
  \bar{Z}_i^{\half}\,,
\eeq
with 
\beq
  Z^{\half}_i\,=\,Z^{L\half}_i\gamma_L+Z^{R\half}_i\gamma_R\,,\hspace{10mm}
  \bar{Z}^{\half}_i\,=\,\bar{Z}^{L\half}_i\gamma_R+\bar{Z}^{R\half}_i\gamma_{L}\,.
\eeq
The fermion propagation amplitude thus is
\beq
  \int d^4 x\,e^{i\,p\cdot x}<\hspace{-1mm}\Omega|T\{\psi_i(x)\bar{\psi}_i(0)\}|
  \Omega\hspace{-1mm}>\,\sim\,
  \frac{\sum_s i<\hspace{-1mm}\Omega|\psi_i(0)|\lambda_{i_s}\hspace{-1mm}>
  <\hspace{-1mm}\lambda_{i_s}|\bar{\psi}_i(0)|\Omega\hspace{-1mm}>}{p^2-m_i^2+i\epsilon}
  \,=\,\frac{i\,Z_i^{\half}\sum_s u^s_i\,\bar{u}^s_i\,\bar{Z}_i^{\half}}
  {p^2-m_i^2+i\epsilon}\,.
\eeq
At tree level $\sum_s u^s_i(p)\,\bar{u}^s_i(p)={\xslash p}+m_i$. When considered radiative corrections Eq.(19) can become \cite{c8}
\beq
  \int d^4 x\,e^{i\,p\cdot x}<\hspace{-1mm}\Omega|T\{\psi_i(x)\bar{\psi}_i(0)\}|
  \Omega\hspace{-1mm}>\,
  \begin{array}{c} \vspace{-7mm} \\ p^2 \rightarrow m_i^2 \\ \vspace{-5mm} \\
  -\hspace{-2mm}-\hspace{-2mm}-\hspace{-2mm}-\hspace{-2mm}-\hspace{-2mm}
  -\hspace{-3mm}\longrightarrow \ear \,
  \frac{i\,Z_i^{\half}(\xslash p+m_i+i x)\bar{Z}_i^{\half}}{p^2-m_i^2+i\epsilon}\,,
\eeq 
where $x$ is a small radiative correction which is proportional to the fermion's decay width at one-loop level \cite{c8}. We note that the result of Eq.(20) is firstly proposed in Ref.\cite{c8} as an assumption, here we derive it under a rational foundation. From Eq.(19), Eq.(16) can also be written as
\beq
  \int d^4 x\,e^{i\,p_j\cdot x}\cdot\cdot\cdot\int d^4 y\,e^{-i\,p_i\cdot y}
  <\hspace{-1mm}\Omega|T\{\psi_j(x)\cdot\cdot\cdot\bar{\psi}_i(y)\}|
  \Omega\hspace{-1mm}>\,\sim\,
  \frac{i\,Z_j^{\half}u_j^s\,\bar{u}^s_j\,\bar{Z}_j^{\half}}{p_j^2-m_j^2+i\epsilon}
  {\cal M}^{amp}(i_s\cdot\cdot\cdot\rightarrow \cdot\cdot\cdot j_s)
  \frac{i\,Z_i^{\half}u_i^s\,\bar{u}^s_i\,\bar{Z}_i^{\half}}{p_i^2-m_i^2+i\epsilon}\,.
\eeq
Thus we obtain from Eqs.(16,17,21) \cite{c8}
\beq
  <\hspace{-1mm}j_s\cdot\cdot\cdot|S|\cdot\cdot\cdot i_s\hspace{-1mm}>\,=\,
  \bar{u}^s_j\,\bar{Z}_j^{\half}{\cal M}^{amp}(i_s\cdot\cdot\cdot\rightarrow \cdot\cdot
  \cdot j_s)Z_i^{\half}u_i^s\,.
\eeq
This is just the usual form of the S-matrix elements containing two external-line fermions \cite{cin1}.

From Eq.(20) the fermion WRC can be obtain by expanding the fermion propagation amplitude at $p^2\rightarrow m_i^2$. Like the boson's case, the fermion WRC becomes \cite{c8}
\beqa
  \bar{Z}_i^{L\half}Z_i^{L\half}&=&(1+\Sigma_{ii}^R(m_i^2))A\,, \nonumber \\
  \bar{Z}_i^{R\half}Z_i^{R\half}&=&(1+\Sigma_{ii}^L(m_i^2))A\,, \nonumber \\
  \bar{Z}_i^{L\half}Z_i^{R\half}&=&Z_i^{L\half}\bar{Z}_i^{R\half}\,,
\eeqa
where the fermion self energy is written as \cite{c8}
\beq
  \Sigma_{ii}(\xslash p)\,=\,\xslash p\gamma_L\Sigma^L_{ii}(p^2)+\xslash p\gamma_R
  \Sigma^R_{ii}(p^2)+m_i\Sigma^{S}_{ii}(p^2)\,,
\eeq
and
\beq
  A\,=\,\left( 1+\left[ \Sigma^L_{ii}+\Sigma^R_{ii}+\Sigma^L_{ii}\Sigma^R_{ii}+
  m_i^2(\Sigma^{L\prime}_{ii}+\Sigma^{R\prime}_{ii}+\Sigma^{L\prime}_{ii}\Sigma^R_{ii}+
  \Sigma^L_{ii}\Sigma^{R\prime}_{ii})+2 m_i(m_i+\delta m_i-m_i\Sigma^S_{ii})
  \Sigma^{S\prime}_{ii} \right]_{p^2=m_i^2} \right)^{-1}\,.
\eeq
On the other hand, from the relationship between the two matrix elements: $<\hspace{-1mm}\lambda_{i_s}|\bar{\psi}_i(0)|\Omega\hspace{-1mm}>\,=\,<\hspace{-1mm}\Omega|\psi_i(0)|\lambda_{i_s}\hspace{-1mm}>^{\dagger}\gamma_0$ we also obtain a hermitian conjugation relationship between the incoming and outgoing fermion WRC (see Eqs.(17,18))
\beq
  \bar{Z}_i^{L\half}\,=\,Z_i^{L\half\ast}\,, \hspace{10mm}
  \bar{Z}_i^{R\half}\,=\,Z_i^{R\half\ast}\,.
\eeq
This coincides with the constraint of Eq.(3) for fermion FRC. But such hermitian conjugation relationship has been broken by the imaginary parts of unstable fermion's propagation amplitudes \cite{c1,c2,c5}. Besides, the result of Eqs.(23) also shows that this hermitian conjugation relationship isn't satisfied by the fermion WRC (see appendix A). 

\section{Determination of the wave-function renormalization constants}

In fact there are also two set WRC for antiparticles. For anti-boson WRC we can introduce them as follows:
\beq
  \int d^4 x\,e^{i\,p_i\cdot x}\cdot\cdot\cdot\int d^4 y\,e^{-i\,p_j\cdot y}
  <\hspace{-1mm}\Omega|T\{\phi_j(y)\cdot\cdot\cdot\phi_i^{\dagger}(x)\}|
  \Omega\hspace{-1mm}>\,\sim\,
  \frac{i<\hspace{-1mm}\Omega|\phi_i^{\dagger}(0)|\lambda_{\bar{i}}\hspace{-1mm}>}
  {p_i^2-m_i^2+i\epsilon}\cdot\cdot\cdot \frac{i<\hspace{-1mm}\lambda_{\bar{j}}|
  \phi_j(0)|\Omega\hspace{-1mm}>}{p_j^2-m_j^2+i\epsilon}
  <\hspace{-1mm}\bar{i}\cdot\cdot\cdot|S|\cdot\cdot\cdot \bar{j}\hspace{-1mm}>\,,
\eeq
where $\lambda_{\bar{i}}$ and $\lambda_{\bar{j}}$ are the outgoing and incoming anti-boson's states of S-matrix elements, and
\beq
  Z_{\bar{i}}^{\half}\,=\,<\hspace{-1mm}\Omega|\phi_i^{\dagger}(0)|
  \lambda_{\bar{i}}\hspace{-1mm}>\,, \hspace{10mm}
  \bar{Z}_{\bar{i}}^{\half}\,=\,<\hspace{-1mm}\lambda_{\bar{i}}|\phi_i(0)|
  \Omega\hspace{-1mm}>\,.
\eeq
The anti-boson's propagation amplitude in interaction vacuum thus is
\beq
  \int d^4 x\,e^{i\,p\cdot x}<\hspace{-1mm}\Omega|T\{\phi_i(0)\phi_i^{\dagger}(x)\}|
  \Omega\hspace{-1mm}>\,\sim\,
  \frac{i<\hspace{-1mm}\Omega|\phi_i^{\dagger}(0)|\lambda_{\bar{i}}\hspace{-1mm}>
  <\hspace{-1mm}\lambda_{\bar{i}}|\phi_i(0)|\Omega\hspace{-1mm}>}{p^2-m_i^2+i\epsilon}
  \,=\,\frac{i\,Z_{\bar{i}}^{\half}\bar{Z}_{\bar{i}}^{\half}}{p^2-m_i^2+i\epsilon}\,.
\eeq
If the charge conjugation is conserved, we will have from Eqs.(9,28)
\beq
  Z_i^{\half}\,=\,Z_{\bar{i}}^{\half}\,, \hspace{10mm}
  \bar{Z}_i^{\half}\,=\,\bar{Z}_{\bar{i}}^{\half}\,.
\eeq
On the other hand, if the charge conjugation isn't conserved, the only variation in standard model is the coupling constants change from real numbers into complex numbers. But in the present problem such variation also disappears, because the Feynman diagrams which generate the matrix elements of Eqs.(9,28) are symmetric in incoming and outgoing states, thus the products of all the coupling constants only include the module squares of the complex coupling constants like the case of the charge conjugation conservation. So Eqs.(30) also holds true in standard model.

Besides, the boson WRC and the anti-boson WRC can also be related by the CPT conservation law. From Eq.(29), Eq.(27) can also be written as
\beq
  \int d^4 x\,e^{i\,p_i\cdot x}\cdot\cdot\cdot\int d^4 y\,e^{-i\,p_j\cdot y}
  <\hspace{-1mm}\Omega|T\{\phi_j(y)\cdot\cdot\cdot\phi_i^{\dagger}(x)\}|
  \Omega\hspace{-1mm}>\,\sim\,
  \frac{i\,Z_{\bar{i}}^{\half}\bar{Z}_{\bar{i}}^{\half}}{p_i^2-m_i^2+i\epsilon}
  \cdot\cdot\cdot \frac{i\,Z_{\bar{j}}^{\half}\bar{Z}_{\bar{j}}^{\half}}
  {p_j^2-m_j^2+i\epsilon}
  {\cal M}^{amp}(\bar{j}\cdot\cdot\cdot\rightarrow \cdot\cdot\cdot \bar{i})\,.
\eeq
From Eqs.(27,28,31) we obtain
\beq
  <\hspace{-1mm}\bar{i}\cdot\cdot\cdot|S|\cdot\cdot\cdot \bar{j}\hspace{-1mm}>\,=\,
  Z_{\bar{j}}^{\half}{\cal M}^{amp}(\bar{j}\cdot\cdot\cdot\rightarrow
  \cdot\cdot\cdot \bar{i})\bar{Z}_{\bar{i}}^{\half}\,.
\eeq
According to the CPT conservation law for bosons
\beq
  <\hspace{-1mm}j\cdot\cdot\cdot|S|\cdot\cdot\cdot i\hspace{-1mm}>\,=\,
  <\hspace{-1mm}\bar{i}\cdot\cdot\cdot|S|\cdot\cdot\cdot \bar{j}\hspace{-1mm}>\,,
\eeq
and the fact ($i$ and $j$ are bosons)
\beq
  {\cal M}^{amp}(i\cdot\cdot\cdot\rightarrow \cdot\cdot\cdot j)\,=\,
  {\cal M}^{amp}(\bar{j}\cdot\cdot\cdot\rightarrow \cdot\cdot\cdot \bar{i})\,,
\eeq
we obtain (see Eqs.(12,32))
\beq
  Z_{\bar{j}}^{\half}\,=\,\bar{Z}_j^{\half}\,, \hspace{10mm}
  \bar{Z}_{\bar{i}}^{\half}\,=\,Z_i^{\half}\,.
\eeq
Since $i$ and $j$ are arbitrary bosons, we finally obtain (see Eqs.(14,30,35))
\beq
  \bar{Z}_{i}\,=\,Z_i\,=\,\bar{Z}_{\bar{i}}\,=\,Z_{\bar{i}}\,=\,
  (1+\Sigma_{ii}^{\prime}(m_i^2))^{-1}\,.
\eeq

Similarly we can get the formula for fermion WRC. For the Green function with two anti-fermion external lines we have
\beq
  \int d^4 x\,e^{i\,p_i\cdot x}\cdot\cdot\cdot\int d^4 y\,e^{-i\,p_j\cdot y}
  <\hspace{-1mm}\Omega|T\{\psi_j(y)\cdot\cdot\cdot\bar{\psi}_i(x)\}|
  \Omega\hspace{-1mm}>\,\sim\,-
  \frac{i<\hspace{-1mm}\lambda_{\bar{j}_s}|\psi_j(0)|\Omega\hspace{-1mm}>}
  {p_j^2-m_j^2+i\epsilon}<\hspace{-1mm}\bar{i}_s\cdot\cdot\cdot|S|
  \cdot\cdot\cdot \bar{j}_s\hspace{-1mm}>\frac{i<\hspace{-1mm}\Omega|\bar{\psi}_i(0)|
  \lambda_{\bar{i}_s}\hspace{-1mm}>}{p_i^2-m_i^2+i\epsilon}\,,
\eeq
where $x^0>y^0$, so the time-ordering operator $T$ exchanges the positions of the fermion fields $\psi_j(y)$ and $\bar{\psi}_i(x)$, thus a minus sign appears. Note that the order of the spins and the gamma matrices keeps unchanged. The anti-fermion WRC can be introduced as follows:
\beq
  <\hspace{-1mm}\Omega|\bar{\psi}_i(0)|\lambda_{\bar{i}_s}\hspace{-1mm}>\,=\,
  \bar{\nu}_i^s\,Z_{\bar{i}}^{\half}\,, \hspace{10mm}
  <\hspace{-1mm}\lambda_{\bar{i}_s}|\psi_i(0)|\Omega\hspace{-1mm}>\,=\,
  \bar{Z}_{\bar{i}}^{\half}\nu_i^s\,,
\eeq
with
\beq
  Z^{\half}_{\bar{i}}\,=\,Z^{L\half}_{\bar{i}}\gamma_R+Z^{R\half}_{\bar{i}}\gamma_L
  \,,\hspace{10mm}
  \bar{Z}^{\half}_{\bar{i}}\,=\,\bar{Z}^{L\half}_{\bar{i}}\gamma_L+
  \bar{Z}^{R\half}_{\bar{i}}\gamma_{R}\,.
\eeq
The anti-fermion propagation amplitude thus is
\beq
  \int d^4 x\,e^{i\,p\cdot x}<\hspace{-1mm}\Omega|T\{\psi_i(0)\bar{\psi}_i(x)\}|
  \Omega\hspace{-1mm}>\,\sim\,-
  \frac{\sum_s i<\hspace{-1mm}\lambda_{\bar{i}_s}|\psi_i(0)|\Omega\hspace{-1mm}>
  <\hspace{-1mm}\Omega|\bar{\psi}_i(0)|\lambda_{\bar{i}_s}\hspace{-1mm}>}
  {p^2-m_i^2+i\epsilon}\,=\, -\frac{i\,\bar{Z}_{\bar{i}}^{\half}\sum_s \nu^s_i\,
  \bar{\nu}^s_i\,Z_{\bar{i}}^{\half}}{p^2-m_i^2+i\epsilon}\,.
\eeq
If the charge conjugation is conserved, according to the formula $u_i^s(p)=-i\gamma_2(\nu_i^s(p))^{\ast}$ we have from Eqs.(17,18)
\beqa
  <\hspace{-1mm}\Omega|\bar{\psi}_i(0)|\lambda_{\bar{i}_s}\hspace{-1mm}>&=&
  -i<\hspace{-1mm}\Omega|\psi_i(0)\gamma_2|\lambda_{i_s}\hspace{-1mm}>^T\gamma_0\,=\,
  -i\left( Z_i^{\half}\gamma_2\,u_i^{s}\right)^T\gamma_0\,=\,
  \bar{\nu}_i^s\left( Z^{L\half}_i\gamma_R+Z^{R\half}_i\gamma_L\right)\,, \nonumber \\
  <\hspace{-1mm}\lambda_{\bar{i}_s}|\psi_i(0)|\Omega\hspace{-1mm}>&=&
  i\,\gamma_0<\hspace{-1mm}\lambda_{i_s}|\gamma_2\,\bar{\psi}_i(0)|
  \Omega\hspace{-1mm}>^T\,=\,i\gamma_0\left( \bar{u}_i^s\gamma_2\,\bar{Z}_i^{\half}
  \right)^T\,=\,\left( \bar{Z}^{L\half}_i\gamma_L+\bar{Z}^{R\half}_i
  \gamma_R \right)\nu_i^s\,,
\eeqa
where the transposition operator $T$ only transposes the spins and the gamma matrices. From Eqs.(38,39) it is obvious that
\beqa
  Z_{\bar{i}}^{L\half}&=&Z_i^{L\half}\,, \hspace{10mm}
  Z_{\bar{i}}^{R\half}\,=\,Z_i^{R\half}\,, \nonumber \\
  \bar{Z}_{\bar{i}}^{L\half}&=&\bar{Z}_i^{L\half}\,, \hspace{10mm}
  \bar{Z}_{\bar{i}}^{R\half}\,=\,\bar{Z}_i^{R\half}\,.
\eeqa
Although in standard model the charge conjugation isn't conserved, Eqs.(41) also keeps unchanged because of the same reason as boson's: the Feynman diagrams which generate the matrix elements of Eqs.(17,38) are symmetric in incoming and outgoing states thus the products of all the coupling constants only include the module squares of the complex coupling constants like the case of the charge conjugation conservation. So Eqs.(42) also hold true in standard model.

Besides, we can also relate the anti-fermion WRC to fermion WRC by the CPT conservation law. From Eq.(40), Eq.(37) can also be written as
\beq
  \int d^4 x\,e^{i\,p_i\cdot x}\cdot\cdot\cdot\int d^4 y\,e^{-i\,p_j\cdot y}
  <\hspace{-1mm}\Omega|T\{\psi_j(y)\cdot\cdot\cdot\bar{\psi}_i(x)\}|
  \Omega\hspace{-1mm}>\,\sim\,-
  \frac{i\,\bar{Z}_{\bar{j}}^{\half}\nu^s_j\,\bar{\nu}^s_j\,Z_{\bar{j}}^{\half}}
  {p_j^2-m_j^2+i\epsilon}{\cal M}^{amp}(\bar{j}_s\cdot\cdot\cdot\rightarrow
  \cdot\cdot\cdot \bar{i}_s) \frac{i\,\bar{Z}_{\bar{i}}^{\half}\nu^s_i\,\bar{\nu}^s_i\,
  Z_{\bar{i}}^{\half}}{p_i^2-m_i^2+i\epsilon}\,.
\eeq
From Eqs.(37,38,43) we obtain
\beq
  <\hspace{-1mm}\bar{i}_s\cdot\cdot\cdot|S|\cdot\cdot\cdot \bar{j}_s\hspace{-1mm}>\,=\,
  \bar{\nu}^s_j\,Z_{\bar{j}}^{\half}{\cal M}^{amp}(\bar{j}_s\cdot\cdot\cdot\rightarrow
  \cdot\cdot\cdot \bar{i}_s)\bar{Z}_{\bar{i}}^{\half}\nu_i^s\,.
\eeq
From the CPT conservation law for fermions
\beq
  <\hspace{-1mm}j_s\cdot\cdot\cdot|S|\cdot\cdot\cdot i_s\hspace{-1mm}>\,=\,-
  <\hspace{-1mm}\bar{i}_s\cdot\cdot\cdot|S|\cdot\cdot\cdot \bar{j}_s\hspace{-1mm}>\,,
\eeq
and Eqs.(22,44) we have
\beq
  \bar{u}^s_j\,\bar{Z}_j^{\half}{\cal M}^{amp}(i_s\cdot\cdot\cdot\rightarrow \cdot\cdot
  \cdot j_s)Z_i^{\half}u_i^s\,=\,-\bar{\nu}^s_j\,Z_{\bar{j}}^{\half}{\cal M}^{amp}
  (\bar{j}_s\cdot\cdot\cdot\rightarrow \cdot\cdot\cdot \bar{i}_s)\bar{Z}_{\bar{i}}^{\half}
  \nu_i^s\,.
\eeq
We can decompose ${\cal M}^{amp}$ into its most general Dirac structure
\beqa
  {\cal M}^{amp}(i_s(p_i)\cdot\cdot\cdot\rightarrow \cdot\cdot\cdot j_s(p_j))&=&
  a(q^2)\,{\xslash q}\,\gamma_L+b(q^2)\,{\xslash q}\,\gamma_R+
  c(q^2)\,{\xslash p}\,\gamma_L+d(q^2)\,{\xslash p}\,\gamma_R+
  e(q^2)\gamma_L+f(q^2)\gamma_R\,, \nonumber \\
  {\cal M}^{amp}(\bar{j}_s(p_j)\cdot\cdot\cdot\rightarrow \cdot\cdot\cdot\bar{i}_s(p_i))
  &=&-a(q^2)\,{\xslash q}\,\gamma_L-b(q^2)\,{\xslash q}\,\gamma_R-
  c(q^2)\,{\xslash p}\,\gamma_L-d(q^2)\,{\xslash p}\,\gamma_R+
  e(q^2)\gamma_L+f(q^2)\gamma_R\,,
\eeqa
with
\beq
  p\,=\,p_i+p_j\,, \hspace{10mm} q\,=\,p_i-p_j\,.
\eeq
Putting Eqs.(47) into Eq.(46) we obtain
\beqa
  &&-a(q^2)(m_j A_L-m_i A_R)Z_i^{L\half}\bar{Z}_j^{L\half}+b(q^2)(m_i A_L-m_j A_R)
  Z_i^{R\half}\bar{Z}_j^{R\half}+c(q^2)(m_j A_L+m_i A_R)Z_i^{L\half}\bar{Z}_j^{L\half}
  \nonumber \\
  &&+d(q^2)(m_i A_L+m_j A_R)Z_i^{R\half}\bar{Z}_j^{R\half}+e(q^2)A_L Z_i^{L\half}
  \bar{Z}_j^{R\half}+f(q^2)A_R Z_i^{R\half}\bar{Z}_j^{L\half} \nonumber \\
  =\hspace{-3mm}&&a(q^2)(m_j B_L-m_i B_R)\bar{Z}_{\bar{i}}^{L\half}Z_{\bar{j}}^{L\half}
  -b(q^2)(m_i B_L-m_j B_R)\bar{Z}_{\bar{i}}^{R\half}Z_{\bar{j}}^{R\half}-c(q^2)
  (m_j B_L+m_i B_R)\bar{Z}_{\bar{i}}^{L\half}Z_{\bar{j}}^{L\half}
  \nonumber \\
  &&-d(q^2)(m_i B_L+m_j B_R)\bar{Z}_{\bar{i}}^{R\half}Z_{\bar{j}}^{R\half}-e(q^2)B_L
  \bar{Z}_{\bar{i}}^{L\half}Z_{\bar{j}}^{R\half}-f(q^2)B_R\bar{Z}_{\bar{i}}^{R\half}
  Z_{\bar{j}}^{L\half}\,,
\eeqa
with
\beqa
  A_L&=&\bar{u}_j^s(p_j)\gamma_L\,u_i^s(p_i)\,, \hspace{10mm}
  A_R\,=\,\bar{u}_j^s(p_j)\gamma_R\,u_i^s(p_i)\,, \nonumber \\
  B_L&=&\bar{\nu}_j^s(p_j)\gamma_L\,\nu_i^s(p_i)\,, \hspace{10mm}
  B_R\,=\,\bar{\nu}_j^s(p_j)\gamma_R\,\nu_i^s(p_i)\,.
\eeqa
Using the fact that
\beq
  A_L\,=\,-B_L\,, \hspace{10mm} A_R\,=\,-B_R\,,
\eeq
we easily obtain from Eq.(49)
\beqa
  Z_i^{L\half}&=&\bar{Z}_{\bar{i}}^{L\half}\,, \hspace{10mm}
  Z_i^{R\half}\,=\,\bar{Z}_{\bar{i}}^{R\half}\,, \nonumber \\
  \bar{Z}_j^{L\half}&=&Z_{\bar{j}}^{L\half}\,, \hspace{10mm}
  \bar{Z}_j^{R\half}\,=\,Z_{\bar{j}}^{R\half}\,.
\eeqa
Combining Eqs.(23,42,52) we finally obtain
\beqa
  \bar{Z}_i^L&=&Z_i^{L}\,=\,\bar{Z}_{\bar{i}}^L\,=\,Z_{\bar{i}}^{L}\,=\,
  (1+\Sigma_{ii}^R(m_i^2))A\,, \nonumber \\
  \bar{Z}_i^R&=&Z_i^{R}\,=\,\bar{Z}_{\bar{i}}^R\,=\,Z_{\bar{i}}^{R}\,=\,
  (1+\Sigma_{ii}^L(m_i^2))A\,.
\eeqa
We note that a similar result is firstly proposed in Ref.\cite{c8} as an assumption. Here we derive it under a rational foundation. 

Now we have totally determined the diagonal boson and fermion WRC. All of the one-loop results of the WRC are listed in the appendix B. There are also off-diagonal WRC, but they are different from the diagonal WRC under the LSZ reduction formula. We note that the off-diagonal WRC should be determined by the prescriptions in Ref.\cite{c2,c8}.

\section{Gauge dependence of physical amplitudes under the conventional and the present wave-function renormalization prescription}

In order to investigate whether the present wave-function renormalization prescription is rational, we calculate two physical processes to see if the physical amplitudes keep gauge invariant under the present wave-function renormalization prescription.

Firstly we discuss the physical process $W^{+}\rightarrow u_i\bar{d}_j$, i.e. the gauge boson $W$ decaying into up-type $i$ and down-type $j$ quarks. At one-loop level Eqs.(53) are equivalent to Eqs.(4.10) of Ref.\cite{c2} if set $\alpha_i=0$. Using the Nielsen identities \cite{c9} Espriu et al. have proved that the physical amplitude ${\cal M}(W^{+}\rightarrow u_i\bar{d}_j)$ is gauge independent under the present wave-function renormalization prescription \cite{c2}.

Secondly we discuss the physical process $Z\rightarrow d_i\bar{d}_i$, i.e. the gauge boson $Z$ decaying into a pair of down-type $i$ quarks. Our numerical calculation has demonstrated the real part of the physical amplitude is gauge-parameter independent, so we only need to discuss the gauge dependence of the imaginary part of the physical amplitude. At one-loop level we have
\beq
  {\cal M}(Z\rightarrow d_i\bar{d}_i)\,=\,\frac{e(2 c_W^2+1)}{12 c_W s_W}\left( \delta Z_Z
  +\delta\bar{Z}^L_{d_i}+\delta\bar{Z}^L_{\bar{d}_i} \right)A_L-\frac{e\,s_W}{6 c_W}
  \left( \delta Z_Z+\delta\bar{Z}^R_{d_i}+\delta\bar{Z}^R_{\bar{d}_i} \right)A_R+
  {\cal M}^{amp}(Z\rightarrow d_i\bar{d}_i)\,,
\eeq
where $e$ is electron charge, $s_W$ and $c_W$ are the sine and cosine of the weak mixing angle, and
\beq
  A_L\,=\,\bar{u}(p_{d_i}){\xslash \epsilon}\gamma_L\nu(p_{\bar{d}_i})\,, \hspace{10mm}
  A_R\,=\,\bar{u}(p_{d_i}){\xslash \epsilon}\gamma_R\nu(p_{\bar{d}_i})\,,
\eeq 
and ${\cal M}^{amp}(Z\rightarrow d_i\bar{d}_i)$ is the amplitude of the amputated diagrams shown in Fig.1.
\begin{figure}[htbp]
\begin{center}
  \epsfig{file=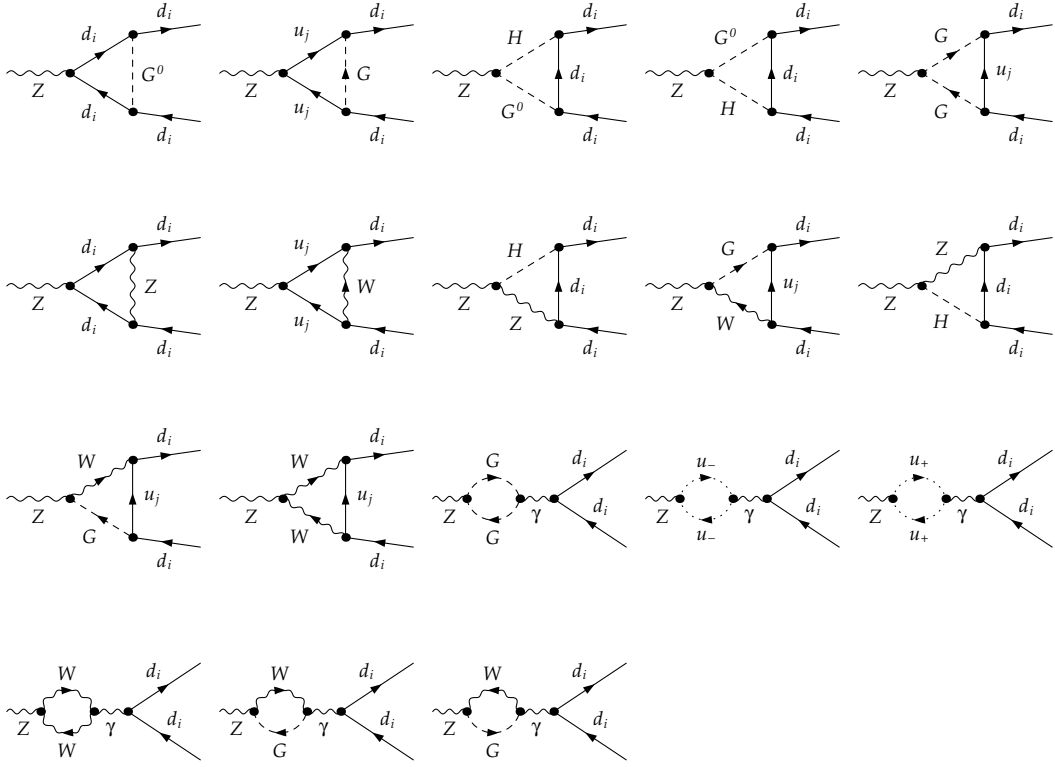,width=14cm} \\
  \caption{One-loop gauge-parameter-dependent diagrams of $Z\rightarrow d_i\bar{d}_i$
  which contain imaginary-part contribution.}
\end{center}
\end{figure}
Using the {\em cutting rules} \cite{c10} we obtain
\beqa
  Im{\cal M}^{amp}(Z\rightarrow d_i\bar{d}_i)|_{\xi}&=&A_L(2 c_W^2+1)\left[
  \frac{e^3}{1152\pi\,c_W^3 s_W^3}(1-4 c_W^2\xi_W)^{3/2}\theta[M_Z-2\sqrt{\xi_W}M_W]
  \right. \nonumber \\
  &-&\frac{e^3}{192\pi\,c_W s_W^3\,x_{d,i}}\sum_j |V_{ji}|^2(x_{d,i}-x_{u,j}-\xi_W)\,B\,
  \theta[m_{d,i}-m_{u,j}-\sqrt{\xi_W}M_W] \nonumber \\
  &-&\left. \frac{e^3}{576\pi\,c_W^3 s_W}
  \left( (\xi_W-1)^2 c_W^4-2(\xi_W-5)c_W^2+1 \right)
  \,C\,\theta[M_Z-M_W-\sqrt{\xi_W}M_W] \right] \nonumber \\
  &+&A_R\left[ -\frac{e^3}{576\pi\,c_W^3 s_W}(1-4 c_W^2\xi_W)^{3/2}
  \theta[M_Z-2\sqrt{\xi_W}M_W] \right. \nonumber \\
  &+&\left. \frac{s_W\,e^3}{288\pi\,c_W^3}
  \left( (\xi_W-1)^2 c_W^4-2(\xi_W-5)c_W^2+1 \right)\,C\,\theta[M_Z-M_W-\sqrt{\xi_W}M_W]
  \right]\,,
\eeqa
where the subscript $\xi$ takes the gauge-parameter-dependent part, $V_{ji}$ is the CKM matrix element \cite{c11}, $M_W$ and $\xi_W$ are the mass and gauge parameter of gauge boson $W$, $M_Z$ is the mass of gauge boson $Z$, $m_{d,i}$ and $m_{u,j}$ are the masses of $d_i$ quark and up-type $j$ quark, and $x_{d,i}=m_{d,i}^2/M_W^2$, $x_{u,j}=m_{u,j}^2/M_W^2$, and
\beq
  B\,=\,\sqrt{\xi_W^2-2(x_{d,i}+x_{u,j})\xi_W+(x_{d,i}-x_{u,j})^2}\,, \hspace{10mm}
  C\,=\,\sqrt{(\xi_W-1)^2c_W^4-2(\xi_W+1)c_W^2+1}\,.
\eeq
We note that the result of Eq.(56) coincides with the results of the conventional loop momentum integral algorithm \cite{c10} and the causal perturbative theory \cite{c12}.

Then we calculate the WRC of gauge boson $Z$ and $d_i$ quark. In Fig.2 we show the one-loop $Z$ self-energy diagrams which are used to calculate the gauge-parameter-dependent imaginary part of $\delta Z_Z$.
\begin{figure}[htbp]
\begin{center}
  \epsfig{file=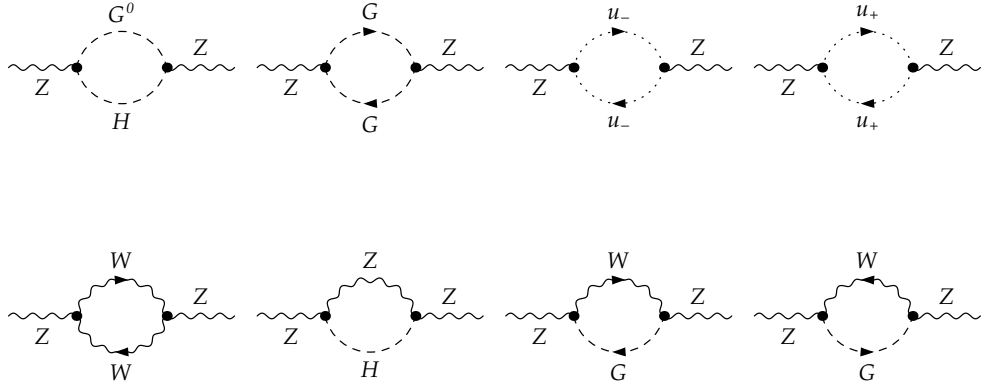,width=13cm} \\
  \caption{One-loop gauge-parameter-dependent $Z$ self-energy diagrams which contain
  imaginary-part contribution.}
\end{center}
\end{figure}
Using Eqs.(36) and the {\em cutting rules} we obtain
\beqa
  Im\,\delta Z_Z |_{\xi}&=&-\frac{e^2}{96\pi\,c_W^2 s_W^2}(1-4 c_W^2\xi_W)^{3/2}
  \theta[M_Z-2\sqrt{\xi_W}M_W] \nonumber \\
  &+&\frac{e^2}{48\pi\,c_W^2}\left( (\xi_W-1)^2 c_W^4-2(\xi_W-5)c_W^2+1 \right)
  \,C\,\theta[M_Z-M_W-\sqrt{\xi_W}M_W]\,.
\eeqa 
The $d_i$ self-energy diagrams used to calculate the gauge-parameter-dependent imaginary part of $d_i$ WRC have been shown in Fig.3. 
\begin{figure}[htbp]
\begin{center}
  \epsfig{file=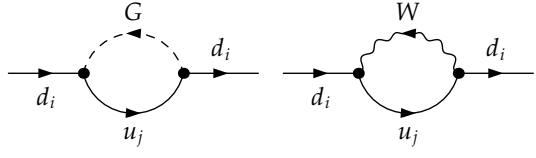,width=7cm} \\
  \caption{One-loop $d_i$ self-energy diagrams which contain imaginary-part contribution.}
\end{center}
\end{figure}
Using Eqs.(53) and the {\em cutting rules} we obtain
\beqa
  Im\,\delta\bar{Z}_{d_i}^R |_{\xi}&=&Im\,\delta\bar{Z}_{\bar{d}_i}^R |_{\xi}\,=\,0\,,
  \nonumber \\
  Im\,\delta\bar{Z}_{d_i}^L |_{\xi}&=&Im\,\delta\bar{Z}_{\bar{d}_i}^L |_{\xi}\,=\,
  \frac{e^2}{32\pi\,s_W^2\,x_{d,i}}\sum_j |V_{ji}|^2
  (x_{d,i}-x_{u,j}-\xi_W)\,B\,\theta[m_{d,i}-m_{u,j}-\sqrt{\xi_W}M_W]\,.
\eeqa
Put Eqs.(56,58,59) into Eq.(54) we finally obtain
\beq
  Im{\cal M}(Z\rightarrow d_i\bar{d}_i)|_{\xi}\,=\,0\,.
\eeq
This means the present wave-function renormalization prescription keeps the physical amplitude of $Z\rightarrow d_i\bar{d}_i$ gauge-parameter independent.

For comparison we evaluate the gauge dependence of physical amplitde $Z\rightarrow d_i\bar{d}_i$ under the conventional wave-function renormalization prescription which discards the imaginary part of unstable particle's WRC \cite{c4}. Under the conventional wave-function renormalization prescription only the last term of the r.h.s. of Eq.(54) contributes to the imaginary part of ${\cal M}(Z\rightarrow d_i\bar{d}_i)$. According to Eqs.(54,56) one readily has
\beqa
  Im{\cal M}(Z\rightarrow d_i\bar{d}_i)|_{\xi}&=&A_L(2 c_W^2+1)\left[
  \frac{e^3}{1152\pi\,c_W^3 s_W^3}(1-4 c_W^2\xi_W)^{3/2}\theta[M_Z-2\sqrt{\xi_W}M_W]
  \right. \nonumber \\
  &-&\frac{e^3}{192\pi\,c_W s_W^3\,x_{d,i}}\sum_j |V_{ji}|^2(x_{d,i}-x_{u,j}-\xi_W)\,B\,
  \theta[m_{d,i}-m_{u,j}-\sqrt{\xi_W}M_W] \nonumber \\
  &-&\left. \frac{e^3}{576\pi\,c_W^3 s_W}
  \left( (\xi_W-1)^2 c_W^4-2(\xi_W-5)c_W^2+1 \right)
  \,C\,\theta[M_Z-M_W-\sqrt{\xi_W}M_W] \right] \nonumber \\
  &+&A_R\left[ -\frac{e^3}{576\pi\,c_W^3 s_W}(1-4 c_W^2\xi_W)^{3/2}
  \theta[M_Z-2\sqrt{\xi_W}M_W] \right. \nonumber \\
  &+&\left. \frac{s_W\,e^3}{288\pi\,c_W^3}
  \left( (\xi_W-1)^2 c_W^4-2(\xi_W-5)c_W^2+1 \right)\,C\,\theta[M_Z-M_W-\sqrt{\xi_W}M_W]
  \right]\,.
\eeqa
This clearly proves that the conventional wave-function renormalization prescription which discards the imaginary part of unstable particle's WRC makes physical amplitude gauge dependent.

Through the two examples we can see the present wave-function renormalization prescription keeps physical amplitude gauge invariant, while other possible wave-function renormalization prescriptions, e.g. the prescriptions in Ref.\cite{c4}, destroys the gauge invariance of physical amplitude. We note that the breaking of the gauge invariance of physical amplitude will break the gauge invariance of physical results.

\section{Conclusion}

We have discussed how to define and totally determine unstable particle's WRC under the postulation of the generalization of the LSZ reduction formula to unstable particles. We introduce two set particle's WRC, and find there are hermitian conjugation relationships between them. But such hermitian conjugation relationships have been broken by the imaginary parts of unstable particle's propagation amplitudes. By introducing two set antiparticle's WRC and the CPT conservation law we find a new wave-function renormalization condition which has been used to totally determine unstable particle's WRC. We have calculated two physical processes to demonstrate the consistence of the present wave-function renormalization prescription with the gauge theory in standard model. We also prove that the conventional wave-function renormalization prescription which discards the imaginary part of unstable particle's WRC makes physical amplitude gauge dependent.

\vspace{5mm} {\bf \Large Acknowledgments} \vspace{2mm}

The author thanks Prof. Xiao-yuan Li for his useful guidance and Prof. Cai-dian Lu for his devoted help.

\vspace{5mm} {\bf \Large Appendix A} \vspace{2mm}

In this appendix we list the evidence that the present wave-function renormalization prescription of Eqs.(23) breaks the hermitian conjugation relationship of Eqs.(26). We calculate the imaginary part of down-type $i$ quark's self energy (the corresponding diagrams have been shown in Fig.3) to demonstrate this problem. At one-loop level Eqs.(23) requires for $d_i$ quark 
\beq
  Im\bar{Z}_{d,i}^{L\half}Z_{d,i}^{L\half}\,=\,\frac{e^2}{32\pi s_W^2 x_{d,i}}
  \sum_j |V_{ji}|^2(x_{d,i}-x_{u,j}-\xi_W)\,B\,\theta[m_{d,i}-m_{u,j}-\sqrt{\xi_W}M_W]\,,
\eeq
with $B$ shown in Eqs.(57). Obviously this result contradicts Eqs.(26).

 \vspace{5mm} {\bf \Large Appendix B} \vspace{2mm}

In this appendix we list all of the one-loop results of boson and fermion WRC. For boson we have from Eqs.(36)
\beq
  \bar{Z}_{i}\,=\,Z_i\,=\,\bar{Z}_{\bar{i}}\,=\,Z_{\bar{i}}\,=\,
  1-\frac{\partial}{\partial p^2}\Sigma_{ii}(m_i^2)\,.
\eeq
For vector boson $\Sigma_{ii}$ is the transverse part of its self energy, i.e. $\Sigma_{ii}^T$ in the following equation:
\beq
  \begin{picture}(72,26)
      \Photon(0,5)(24,5){2}{4}
      \GCirc(36,5){12}{0.5}
      \Photon(48,5)(72,5){2}{4}
      \Text(10,13)[]{$i,\mu$}
      \Text(10,-3)[]{$k$}
      \Text(62,13)[]{$i,\nu$}
  \end{picture}\,=\,-i g^{\mu\nu}(k^2-m_i^2)-i(g^{\mu\nu}-\frac{k^{\mu}k^{\nu}}{k^2})
  \Sigma^T_{ii}(k^2)-i\frac{k^{\mu}k^{\nu}}{k^2}\Sigma^L_{ii}(k^2)\,.
\eeq
For fermion we have from Eqs.(53)
\beqa
  \bar{Z}_i^L&=&Z_i^{L}\,=\,\bar{Z}_{\bar{i}}^L\,=\,Z_{\bar{i}}^{L}\,=\,
  1-\Sigma^L_{ii}(m_i^2)-m_i^2\frac{\partial}{\partial p^2}\left(\Sigma^L_{ii}(p^2)+
  \Sigma^R_{ii}(p^2)+2\Sigma^S_{ii}(p^2)\right)_{p^2=m_i^2}\,, \nonumber \\
  \bar{Z}_i^R&=&Z_i^{R}\,=\,\bar{Z}_{\bar{i}}^R\,=\,Z_{\bar{i}}^{R}\,=\,
  1-\Sigma^R_{ii}(m_i^2)-m_i^2\frac{\partial}{\partial p^2}\left(\Sigma^L_{ii}(p^2)+
  \Sigma^R_{ii}(p^2)+2\Sigma^S_{ii}(p^2)\right)_{p^2=m_i^2}\,.
\eeqa


\begin{thebibliography}{99}
\bibitem{cin1}
We note that because of the difference between Eq.(16) and Eq.(10) of Ref.[9] one cannot directly obtain the result of Eq.(22) from Ref.[9] (note that the fermion spinors $\bar{u}(p_1)$ and $u(k_1)$ in Eq.(12) of Ref.[9] are added by hand).
\end{thebibliography}
\end{document}